\documentclass[12pt,pra,aps,amssymb,amsfonts,amsmath,tightenlines]{revtex4}

\usepackage[dvips]{graphicx}
\usepackage{dsfont, color}

\newcommand{\re}{\mbox{$\rm e$}}
\newcommand{\rd}{\mbox{$\rm d$}}
\newcommand{\half}{\mbox{$\textstyle \frac{1}{2}$}}
\DeclareMathOperator{\sgn}{sgn}

\begin{document}
\title{A game of information}
\author{Dorje C. Brody}

\affiliation{School of Mathematics and Physics, 
University of Surrey, Guildford GU2 7XH, UK, and \\ 
Department of Mathematics, Imperial College London, 
London SW7 2BZ, UK}

\date{\today}

\begin{abstract}
A game of information concerns two players transmitting messages that are obscured by noise. A receiver digests the combination of the two information sources and makes an assessment rationally. The aim of the players is to generate opposing assessments for the receiver by choosing optimal signal-to-noise ratios of their information. It is shown that this problem can be reduced into an elementary infinite game on the square, thus admitting a complete equilibrium solution. The problem is extended to allow for one player to disseminate disinformation, for which the optimal strategies become more intricate. Further generalisations of the game are proposed. 
\end{abstract}

\maketitle

%%%%%%%%%%%%

\section{Introduction}
\label{sec:1}

In today's society endowed with a ``radically new technology of information 
dissemination through social media'' \cite{Rycroft}, achieving an objective 
that involves a shift in public opinion will have to rely in some ways on a game 
of information. To explain what I mean by a ``game of 
information'' let me consider a simple example involving two players, A and B. 
Both players reveal noisy information about a binary topic of interest to the public. 
The latter digest the combination of two information sources rationally (that is, 
following the Bayes rule) and make an inference on the topic. The two players must 
preselect the flow rates (signal-to-noise ratios) of their information, subject to 
budget constraints, so as to achieve their opposing objectives: player A wishes 
to sway the public in one direction, while player B wishes to sway them in the 
other direction, on a specific future date. However, they do not know which 
choice the other player will make. The question then is to determine optimal 
strategies for the players to choose their information flow rates. The main 
purpose of the present paper is to show, in the case of time-independent 
information flow rates, that the problem can be translated into a standard 
infinite game on the unit square \cite{Karlin}, thus admitting a Nash equilibrium 
for any parameter choice.  

It should be evident that the game of information constructed here, and its 
generalisations, have many practical applications. Perhaps the most apparent 
example is a two-candidate electoral competition \cite{Brody1}. 
The two players are the two campaign teams, and the 
objective is to maximise the public support for their candidates on the future 
election day. Another application is in marketing whereby two leading companies 
of a product wish to maximise their market shares over a given financial period, 
and wish to decide on optimal advertisement strategies. Yet another example 
concerns a situation such as player A representing climate scientists and player 
B representing fossil fuel industry, and they wish to influence public opinion on 
climate policies over a given time period. 

It is often the case that information is contaminated with disinformation, for 
example, player A plays a fair game but player B cheats by spreading 
disinformation. Here, 
by disinformation I mean a false information that is disseminated with an 
intension of misguiding the public \cite{BM}. Hence genuine rumours, 
speculations, and misunderstandings are counted as noise, not disinformation. 
In this respect, the third example mentioned above fits well with the setup: 
Climate forecasting 
is prone to errors, but these errors can be viewed as noise, whereas messages 
from climate sceptics are often contaminated with disinformation. My main focus 
here is to introduce a game of information in the context of genuine ``fair'' 
information, free of disinformation, and show how in a simple situation it can be 
reduced to an elementary exercise in game theory. Nevertheless, I shall pay 
attention to the effect of disinformation and show how the presence of 
disinformation can complicate the strategies. 

With this in mind, the present paper is organised as follows. In 
Section~\ref{sec:2}, I briefly review how different information sources can be 
combined, and how they interact for a receiver of two information sources. I 
then define a concrete example of a game of information involving two players. 
Then in Section~\ref{sec:3}, I show how this information game can be reduced 
to a simple infinite game on the unit square, and work out the solution in closed 
form. In particular, one finds that depending on the value of the correlation of the 
noise terms in the two information sources, there is either a pure strategy Nash 
equilibrium or a mixed strategy. In Section~\ref{sec:4}, I introduce a disinformation 
term in one of the players and show how the game is modified. A full analytic 
solution to the game of disinformation is then worked out in Section~\ref{sec:5}, 
showing that in some parameter ranges a mixed Nash equilibrium exists. I conclude in 
Section~\ref{sec:6} with two other open challenges in the game of information.

\section{Information fusion and the statement of the problem}
\label{sec:2}

In communication theory, transmission of information through a channel is 
modelled by a combination of a signal term, a noise term, and a signal-to-noise 
ratio. Noise that obscures the signal is typically modelled by an additive Gaussian 
process. The signal can itself be a time series, but here I consider a simpler 
case where it is given by a fixed (in time) binary random variable $X$. Without 
loss of generality we can assume that $X$ takes the value $0$ with the 
\textit{a priori} probability $p$ and the value $1$ with the \textit{a priori} 
probability $1-p$. The \textit{a priori} probabilities can be viewed as the 
current proportion of rational population who would choose the alternatives 
labelled by the two values that $X$ can take. If one assumes, as is commonly 
done, that signal contamination is modelled by an additive Gaussian noise 
$\epsilon_t$, and if we let $\sigma$ be the (time-independent) signal-to-noise 
ratio, then information reaching the receiver is given by the time series 
$\xi_t=\sigma X t + \epsilon_t$, where $X$ and $\epsilon_t$ are assumed 
independent. 

In the present consideration we have two players, A and B, both disseminating 
information about $X$ in a noisy environment. Writing $\sigma_A$ and 
$\sigma_B$ for the respective information flow rates, a receiver (public) obtains 
the combination of 
\begin{eqnarray}
\xi_t^A=\sigma_A X t + \epsilon_t^A \quad {\rm and} \quad 
\xi_t^B=\sigma_B X t + \epsilon_t^B \, , 
\label{eq:1}
\end{eqnarray} 
where the two noise terms $\epsilon_t^A$ and $\epsilon_t^B$ will be modelled 
by standard Brownian motions, with correlation $\rho\in[-1,1]$. It is known 
\cite{BY2} that when two information sources of the form (\ref{eq:1}) for the 
same signal $X$ are available, they can be combined into a single effective 
information 
\begin{eqnarray}
\xi_t=\sigma X t + \epsilon_t \, , 
\label{eq:2}
\end{eqnarray} 
where 
\begin{eqnarray}
\sigma = \sqrt{\frac{\sigma_A^2 - 2\rho\sigma_A \sigma_B + \sigma_B^2}{1-\rho^2}}
\label{eq:3}
\end{eqnarray} 
is the effective information flow rate, and 
\begin{eqnarray}
\epsilon_t = \frac{\sigma_A-\rho\sigma_B}{\sigma(1-\rho^2)} \, \epsilon_t^A + 
\frac{\sigma_B-\rho\sigma_A}{\sigma(1-\rho^2)} \, \epsilon_t^B
\label{eq:4}
\end{eqnarray} 
is the effective noise. 

When a rational receiver detects the combined information (\ref{eq:3}), the 
prior probability $p={\mathbb P}(X=0)$ will be updated to the posterior 
probability $\pi_t={\mathbb P}(X=0|\{\xi_t\})$ according to the Bayes rule, 
which in the present example is given by 
\begin{eqnarray}
\pi_t = \frac{p}{p+(1-p)\re^{\sigma \xi_t - \frac{1}{2}\sigma^2 t}} \, . 
\label{eq:5} 
\end{eqnarray} 
In the theory of signal detection, this is known as the Wonham filter 
\cite{wonham}. 
In a Gibbsian ensemble interpretation, we can regard the posterior probability 
$\pi_t$ as the proportion of the population who would make the choice $X=0$ 
at time $t$, based on the prior proportion $p$ at time zero, and on the 
information strategy (\ref{eq:1}). As an example system modelled by this 
framework, consider a two candidate electoral competition. In this case, 
$\{\xi_t^A\}$ represents the information strategy of the campaign team of 
candidate A, and similarly $\{\xi_t^B\}$ for candidate B. If the election were to 
take place at $t=T$, then the likelihood of candidate A winning the future 
election, as of time zero, given the informational strategy (\ref{eq:1}), is 
given by ${\mathbb P}(\pi_T>\frac{1}{2})$. Player (candidate) A therefor 
wishes to maximise ${\mathbb P}(\pi_T>\frac{1}{2})$, while player B 
wishes to minimise it. The game of information in this example is then for 
the two players to choose, at time zero, the information flow rates 
$\sigma_A$ and $\sigma_B$ so as to achieve their objectives, without the 
knowledge of the choice of the other player. Needless to add, disseminating 
information (e.g., advertisement) is costly, so it will be desirable to impose 
budget constraints, which, for simplicity, will be imposed in the form 
$\sigma_A\leq1$ and $\sigma_B\leq1$.

\section{Solution to the simple information game}
\label{sec:3}

The solution to the simple game of information specified in the previous section 
can be summarised as follows. 

\vspace{0.2cm}
\noindent 
{\bf Proposition 1}. \textit{
Assuming that $p>1/2$, 
the Nash equilibrium solution to the game of information takes 
the following form. {\rm(i)} For $-1<\rho<0$ the optimal pure strategies are 
$\sigma_A^*=0$ and $\sigma_B^*=1$; {\rm(ii)} For $0\leq\rho<1/\sqrt{2}$ the 
optimal pure strategies are $\sigma_A^*=\rho$ and $\sigma_B^*=1$; {\rm(iii)} 
For $1/\sqrt{2}\leq\rho<1$ the optimal mixed strategies are 
$\sigma_A^*=1/2\rho$, while $\sigma_B^*=0$ with probability $1-1/2\rho^2$ 
and $\sigma_B^*=1$ with probability $1/2\rho^2$. When $p<1/2$, the 
strategies of the two players are flipped around.}
\vspace{0.2cm}

\noindent 
{\it Proof}. From (\ref{eq:5}) we find, on account of the monotonicity of $\pi_T$ 
on $\xi_T$, that the condition $\pi_T>K$ for any $K\in
[0,1]$ is equivalent to the condition that 
\begin{eqnarray}
\xi_T < \frac{ \log\left[ \frac{p(1-K)}{K(1-p)}\right] + \frac{1}{2} \sigma^2 T}
{\sigma} \equiv \xi^* \, .  
\end{eqnarray} 
Because the density function for the sum of two independent random variables 
$\sigma X T$ and $\epsilon_T$ is just the convolution of the densities of 
individual random variables, we find that the density function for 
$\xi_T=\sigma X T + \epsilon_T$ is given by the convolution 
\begin{eqnarray}
\rho(\xi) = \frac{1}{\sqrt{2\pi T}} 
\left[ p\, \re^{-\frac{1}{2T}\xi^2} + (1-p)\, 
\re^{-\frac{1}{2T}(\xi-\sigma T)^2} \right] \, . 
\end{eqnarray}
A calculation then shows \cite{Brody1,Brody2} that for any $K\in[0,1]$ we have  
\begin{eqnarray}
{\mathbb P}\left( \pi_T> K \right) = \int_{-\infty}^{\xi^*} \rho(\xi) \, \rd\xi 
= p\, N(d^+) + (1-p)\, N(d^-) \, , 
\label{eq:8} 
\end{eqnarray}
where 
\begin{eqnarray}
N(x) = \frac{1}{2\pi} \int_{-\infty}^x \re^{-\frac{1}{2}z^2} \, \rd z 
\label{eq:9} 
\end{eqnarray}
is the cumulative normal distribution function, and   
\begin{eqnarray}
d^\pm =  \frac{ \log\left[ \frac{p(1-K)}{K(1-p)}\right] \pm \frac{1}{2} \sigma^2 T} 
{\sigma\sqrt{T}} \, .
\label{eq:10} 
\end{eqnarray}
Recall that the objective of player A is to maximise 
${\mathbb P}\left( \pi_T> K \right)$ and the objective of player B is to minimise 
${\mathbb P}\left( \pi_T> K \right)$.  It is straightforward to check that the 
expression for ${\mathbb P}\left( \pi_T> K \right)$ in 
(\ref{eq:8}), as a function of $\sigma$, which is the only variable from which 
the dependencies on $\sigma_A$ and $\sigma_B$ enter through (\ref{eq:3}), is 
monotonically increasing in $\sigma$ if $p<K$ and monotonically decreasing 
in $\sigma$ if $p>K$. This means that if $p<K$, player A wishes to maximise 
$\sigma$ and player B wishes to minimise $\sigma$, whereas if $p>K$, player 
A wishes to minimise $\sigma$ and player B wishes to maximise $\sigma$; in 
both cases, under the constraints $\sigma_A\leq1$ and $\sigma_B\leq1$. 
Consider the latter case $p>K$ so that the objective is for player A to minimise 
and player B to maximise $\sigma$ under the constraints. 

The game of information has therefore been transformed into 
an elementary continuous two-player infinite game on the unit square 
\cite{Karlin}. In particular, we have 
\begin{eqnarray}
\frac{\partial^2\sigma}{\partial \sigma_A^2} = 
\frac{\sigma_B^2}{\sigma^3(1-\rho^2)}\geq 0 
\quad {\rm and} \quad 
\frac{\partial^2\sigma}{\partial \sigma_B^2} = 
\frac{\sigma_A^2}{\sigma^3(1-\rho^2)}\geq 0 \, ,
\label{eq:11}
\end{eqnarray}
hence $\sigma$ is a convex function of the two variables. Consider first the case 
(i) where $-1<\rho\leq0$. Because the maximum value of a convex function over 
a finite interval $[0,1]$ is necessarily located at the boundary, player B who wishes 
to maximise $\sigma$ will choose either $\sigma_B=0$ or $\sigma_B=1$. In either 
case, $\sigma|_{\sigma_B=0}=\sigma_A/\sqrt{1-\rho^2}$ and 
$\sigma|_{\sigma_B=1}=\sqrt{(\sigma_A^2-2\rho\sigma_A+1)/(1-\rho^2)}$ for 
$\rho<0$ are both increasing 
in $\sigma_A$, so player A who wishes to minimise $\sigma$ will 
choose the smallest possible value $\sigma_A^*=0$. Therefore, to maximise 
$\sigma|_{\sigma_A=0}=\sigma_B/\sqrt{1-\rho^2}$, player B must choose 
$\sigma_B^*=1$. 

Next, consider the case (ii) where $0<\rho<1/\sqrt{2}$. In this case, on account 
of convexity player $B$ will again choose one of the boundary values 
$\sigma_B=0$ or $\sigma_B=1$. If player B chooses $\sigma_B=0$, then to 
minimise $\sigma$, player A will choose $\sigma_A=0$, but then we have 
$\sigma=0$, which is suboptimal for player B so player B will not choose 
$\sigma_B=0$. Instead, player B chooses $\sigma_B^*=1$. In this case, to 
minimise $\sigma|_{\sigma_B=1}$, player A will choose $\sigma_A^*=\rho$. 

Finally, in case (iii) where $1/\sqrt{2}\leq\rho<1$, a mixed strategy becomes feasible. 
Player B, the maximiser, continues to choose either $\sigma_B=0$ or 
$\sigma_B=1$, but this time with probabilities $r$ and $1-r$, respectively. Player 
A therefore must find a strategy that is indifferent to the random choice made by 
player B. That is, the choice $\sigma_A$ of player A must satisfy 
\begin{eqnarray}
\sqrt{\frac{\sigma_A^2}{1-\rho^2}} = 
\sqrt{\frac{\sigma_A^2 - 2\rho\sigma_A  + 1}{1-\rho^2}} \, ,
\label{eq:12}
\end{eqnarray} 
the solution to which gives $\sigma_A^*=1/2\rho$. Now we observe that the 
minimiser of $\sigma$ also minimises its expectation 
\begin{eqnarray}
{\mathbb E}[\sigma] = r \sqrt{\frac{\sigma_A^2}{1-\rho^2}} + (1-r) 
\sqrt{\frac{\sigma_A^2 - 2\rho\sigma_A  + 1}{1-\rho^2}} 
\label{eq:13}
\end{eqnarray}
over the random strategy of player B. Differentiating this with respect to 
$\sigma_A$ and setting the result to zero, we obtain 
\begin{eqnarray}
 r + (1-r) \frac{\sigma_A-\rho}{\sqrt{\sigma_A^2 - 2\rho\sigma_A  + 1}} = 0 \, .  
\label{eq:14}
\end{eqnarray}
Because (\ref{eq:14}) must hold for $\sigma_A=\sigma_A^*=1/2\rho$, solving 
it for $r$ we deduce that $r^*=1-1/2\rho^2$. A probabilistic strategy, however, 
is available to player B only if $r^*\geq0$, that is, only if $\rho\geq1/\sqrt{2}$. 
Hence for $0<\rho<1/\sqrt{2}$ the strategy reduces to a pure strategy. Finally, 
we remark that if $p<K$, player A becomes maximiser and player B 
becomes minimiser, but $\sigma$ is symmetric in $\sigma_A$ and $\sigma_B$, 
so all the results are flipped. 
\hfill $\Box$

It is worth noting that the actual gain, or loss, of the two players is not the change 
in the value of $\sigma$, but change in the probability ${\mathbb P}\left( 
\pi_T> \frac{1}{2} \right)$. As a function of the prior $p$, however, 
${\mathbb P}\left( \pi_T> \frac{1}{2} \right)$ is antisymmetric about $p=\frac{1}{2}$. 
Hence any gain in the probability for one player matches exactly the loss for the 
other player, so this version of information game is indeed a zero-sum game.

\section{Information game with disinformation} 
\label{sec:4} 

It is often the case in practice that some players tend to cheat when playing games. 
In a game of information this means an act of disseminating disinformation. 
Disinformation is defined here to be a bias in the noise that is intended to misguide 
the receiver of the information \cite{BM}. Hence rumours, speculations, or genuine 
misunderstandings that possess no intent will count as noise and not disinformation. 
Here we consider the case whereby player B releases information that is 
contaminated with disinformation, released at constant rate $\mu$. Hence we 
have the information strategies 
\begin{eqnarray}
\xi_t^A=\sigma_A X t + \epsilon_t^A \quad {\rm and} \quad 
\xi_t^B=\sigma_B X t + \epsilon_t^B + \mu t \, .
\label{eq:15}
\end{eqnarray} 
The assumption is that player A knows the existence of the drift term in the 
information transmitted from player B, but not the value of $\mu$. The receiver 
of the combined information (the public), on the other hand, are unaware of the 
existence of $\mu$, and therefore are misguided as a result. This is indeed 
not an uncommon situation --- take, for example, a ``game'' between those 
who disseminate online disinformation with an intent to generate violent social 
unrest and a government trying to restore order and calm --- making the game 
worthwhile analysing. 

To investigate the effect of disinformation, it is an easy exercise to check 
\cite{BY2} that the combined information (\ref{eq:2}) can alternatively be 
expressed in the form 
\begin{eqnarray}
\xi_t = \frac{\sigma_A-\rho\sigma_B}{\sigma(1-\rho^2)} \, \xi_t^A + 
\frac{\sigma_B-\rho\sigma_A}{\sigma(1-\rho^2)} \, \xi_t^B \, . 
\label{eq:15}
\end{eqnarray}
It follows that the existence of disinformation in $\xi_t^A$ induces the 
following bias 
\begin{eqnarray}
\xi_t \to \xi_t + \frac{(\sigma_B-\rho\sigma_A)}{\sigma(1-\rho^2)} \, \mu t \, . 
\label{eq:16}
\end{eqnarray}
Note that the \textit{intention} of player B in choosing a positive bias is to enhance 
the probability for the choice $X=1$, whereas if $\mu<0$ then the intension is to 
enhance the probability for the choice $X=0$. However, because two information 
sources interact in a nontrivial manner to reach the receiver, if $\rho>0$, then 
(\ref{eq:16}) shows that the intended effect can be reversed if $\sigma_B<
\rho\sigma_A$. In other words, player A can reverse the effect of disinformation 
simply by increasing their information flow rate \cite{BY2}. This observation makes 
the game somewhat nontrivial, because there can be counterintuitive situations. 

The fact that receivers of the combined information are unaware of the existence 
of $\mu$ (otherwise the effect of $\mu$ will be disregarded) 
means that they will follow exactly the same analysis leading to 
(\ref{eq:8}), except that their assessments will be skewed by the shift 
\begin{eqnarray}
d^\pm \to  \frac{ \log\left[ \frac{p(1-K)}{K(1-p)}\right] \pm \frac{1}{2} \sigma^2 T 
- \frac{\sigma_B-\rho\sigma_A}{1-\rho^2}\mu T}{\sigma\sqrt{T}} \, .
\label{eq:18} 
\end{eqnarray}
With this in mind, a ``game of disinformation'' can be stated as follows. The 
objective of player A is to maximise, and that of player B is to minimise 
${\mathbb P}\left(\pi_T> K \right)$ of (\ref{eq:8}), with $d^\pm$ given by 
(\ref{eq:18}). Player A must choose $\sigma_A$, and player B must choose 
both $\sigma_B$ and $\mu$, to meet their objectives, subject to the budget 
constraints $\sigma_A\leq1$, $\sigma_B\leq1$, and $|\mu|\leq1$.

\section{Analysis of disinformation game}
\label{sec:5} 

The existence of the drift $\mu$ breaks the symmetry so that there is no 
simple solution to the game of disinformation. Nevertheless, an analysis yields 
the following conclusion. 

\vspace{0.2cm}
\noindent 
{\bf Proposition 2}. \textit{
{\rm(i)} For $-1<\rho<0$ the optimal pure strategies are $\mu^*=1$, 
$\sigma_A^*=0$, and $\sigma_B^*=1$; {\rm(ii)} For $0\leq\rho$ the optimal 
strategy for the disinformation is $\mu^*=\sgn(\sigma_B-\rho\sigma_A)$. 
The optimal mixed strategy for the information flow rates are 
$\sigma_A^*=1/2\rho$, while $\sigma_B^*=0$ with probability 
\begin{eqnarray}
r^* = \frac{1}{2\rho^2} \left[ \frac{(2\rho^2-1)\left[ 
\frac{2\rho^2(1-\rho^2)}{T} \log\left( \frac{p}{1-p}\right) - \rho^2 
- \frac{1}{4} \tanh\left( \frac{T}{2(1-\rho^2)}\right) \right]+ \rho^2} 
{\frac{2\rho^2(1-\rho^2)}{T} \log\left( \frac{p}{1-p}\right) - \rho^2 
- \frac{1}{4} \tanh\left( \frac{T}{2(1-\rho^2)}\right) + 1}  \right]  
\label{eq:z32}
\end{eqnarray} 
and $\sigma_B^*=1$ with probability $1-r^*$, provided that 
if 
$\rho\geq1/\sqrt{2}$ and either 
\begin{eqnarray} 
p \geq \frac{1}{1+\re^{-\frac{T}{2\rho^2(1-\rho^2)} \left( \rho^2-1+\frac{1}{4}\tanh 
\left(\frac{T}{2(1-\rho^2)} \right) \right)} } 
\label{eq:z20}
\end{eqnarray}
or 
\begin{eqnarray} 
p \leq \frac{1}{ 1+\re^{-\frac{T}{2\rho^2(1-\rho^2)} \left( 
\frac{2\rho^2(\rho^2-1)}{2\rho^2-1}+\frac{1}{4}\tanh 
\left(\frac{T}{2(1-\rho^2)} \right) \right)} }  
\label{eq:z21}
\end{eqnarray}
holds, or if $\rho<1/\sqrt{2}$ and 
\begin{eqnarray} 
\frac{1}{1+\re^{-\frac{T}{2\rho^2(1-\rho^2)} \left( \rho^2-1+\frac{1}{4}\tanh 
\left(\frac{T}{2(1-\rho^2)} \right) \right)} } 
< p < 
\frac{1}{ 1+\re^{-\frac{T}{2\rho^2(1-\rho^2)} \left( 
\frac{2\rho^2(\rho^2-1)}{2\rho^2-1}+\frac{1}{4}\tanh 
\left(\frac{T}{2(1-\rho^2)} \right) \right)} }  
\label{eq:z22}
\end{eqnarray}
holds. Otherwise, the optimal pure strategy is $\sigma_A^*=1$ and $\sigma_B^*=1$.}
\vspace{0.2cm}

\noindent 
{\it Proof}. 
Consider first the case $\rho<0$. In this case, $\sigma_B-\rho \sigma_A>0$, so 
the best strategy for player B, for the parameter $\mu$, is to set $\mu^*=1$. For 
player A, on the other hand, the disadvantage of having nonzero $\mu$ can be 
minimised by choosing $\sigma_A=0$. This choice also minimises $\sigma$, 
which, in the absence of $\mu$ will be optimal. Hence the optimal strategy for 
player A is to set $\sigma_A^*=0$. In this case, player B wishes to minimise 
${\mathbb P}\left(\pi_T> 1/2 \right)$. By setting $\sigma_B^*=1$, it simultaneously 
maximises the impact of $\mu$ and maximises $\sigma$, which is optimal 
when $p\geq/2$. Indeed, making use of the relation 
\begin{eqnarray}
\frac{\sigma_B-\rho \sigma_A}{\sigma(1-\rho^2)}  = 
\sqrt{\sigma^2-\sigma_A^2} 
\end{eqnarray}
one sees that for $\mu=1$ the function ${\mathbb P}\left(\pi_T> 1/2 \right)$ is 
monotonically decreasing in both $\sigma_A$ and $\sigma_B$, thus justifying 
$\sigma_A^*=0$ and $\sigma_B^*=1$. 

For $\rho\geq0$ the analysis becomes intricate. Player B for sure wishes to 
maximise the impact of disinformation, and this is achieved by setting 
$\mu^*=\sgn(\sigma_B-\rho\sigma_A)$. To proceed, it will be convenient to 
change the variable $\sigma_B\to u_B$ according to 
\begin{eqnarray}
u_B = \frac{|\sigma_B-\rho\sigma_A|}{\sqrt{1-\rho^2}} \, . 
\label{eq:20}
\end{eqnarray} 
Then we have $\sigma^2=u_B^2+\sigma_A^2$. For simplicity of notation define 
$\Lambda = \log[p/(1-p)]$ and $\alpha=T/\sqrt{1-\rho^2}$. Then we have 
\begin{eqnarray}
d^\pm = \frac{\Lambda\pm\frac{1}{2}(u_B^2+\sigma_A^2)T-\alpha u_B}
{\sqrt{(u_B^2+\sigma_A^2)T}} \, . 
\label{eq:21}
\end{eqnarray}

Player B seeks to find $u_B$ that minimises ${\mathbb P}\left(\pi_T> 1/2 \right)$. 
By differentiation, making use of the identity $(1-p)\,n(d^-) = 
p\,n(d^+)\,\re^{-\alpha u_B}$ valid for the standard normal density function $n(x)$, 
and equating the result to zero, we find 
\begin{eqnarray}
\frac{2(\Lambda u_B+\alpha \sigma_A^2)}{u_BT(u_B^2+\sigma_A^2)} = 
\tanh\left( \frac{\alpha u_B}{2}\right) \, . 
\label{eq:22}
\end{eqnarray}
Player A, on the other hand, seeks to maximise ${\mathbb P}\left(\pi_T> 1/2 
\right)$. In view of the envelope theorem, taking the partial derivative 
with respect to $\sigma_A$ and setting the result to zero, we find the relation 
\begin{eqnarray}
\frac{2(\Lambda -\alpha u_B)}{T(u_B^2+\sigma_A^2)} = 
\tanh\left( \frac{\alpha u_B}{2}\right) \, . 
\label{eq:23}
\end{eqnarray}
Comparing (\ref{eq:22}) and (\ref{eq:23}) we find that if there is a simultaneous 
Nash equilibrium solution then it has to be that $u_B^2=-\sigma_A^2$, which 
is not possible unless $\sigma_A=\sigma_B=0$. It follows that there is no such 
equilibrium solution in the interior region $\sigma_A>0$ and $u_B>0$. The 
solution $u_B=0$, however, is not optimal for player B because the derivative 
of ${\mathbb P}\left(\pi_T> 1/2 \right)$ at $u\to0$ is negative. Player B therefore 
seeks to move as far away from $u_B=0$ as possible, and this implies that 
possible solutions are either $\sigma_B^*=0$ or $\sigma_B^*=1$. 

This leads to a mixed strategy whereby player B chooses $\sigma_B^*=0$ 
with probability $r$ and $\sigma_B^*=1$ with probability $1-r$. Player A thus 
seeks an indifferent $\sigma_A=\sigma_A^*$ such that 
\begin{eqnarray} 
\big[ p\, N(d^+) + (1-p)\, N(d^-) \big]_{\sigma_B=0} = 
\big[ p\, N(d^+) + (1-p)\, N(d^-) \big]_{\sigma_B=1} \, ,
\end{eqnarray}
which is possible only if $d^{\pm}|_{\sigma_B=0}=d^{\pm}|_{\sigma_B=1}$. 
Solving this for $\sigma_A$ we find that $\sigma_A^* = 1/2\rho$. The constraint 
$\sigma_A\leq1$ however shows that a mixed strategy is possible only if 
$\rho\geq1/2$. Hence in this range of correlation, player B plays a mixed 
strategy so that the expected probability for the event $\pi_T>1/2$ is 
\begin{eqnarray}
P = r \big[ p\, N(d^+) + (1-p)\, N(d^-) \big]_{\sigma_B=0}  + (1-r) 
\big[ p\, N(d^+) + (1-p)\, N(d^-) \big]_{\sigma_B=1} \, . 
\end{eqnarray}
We thus consider the condition 
\begin{eqnarray}
\frac{\partial P}{\partial \sigma_A} = \frac{\partial \sigma^2}{\partial \sigma_A} 
\frac{\partial P}{\partial \sigma^2} +  \frac{\partial u_B}{\partial \sigma_A} 
\frac{\partial P}{\partial u_B} = 0 \, , 
\end{eqnarray} 
which must hold when $\sigma_A=1/2\rho$. In other words, we require 
\begin{eqnarray}
r \left[ \frac{\partial \sigma^2}{\partial \sigma_A} 
\frac{\partial P}{\partial \sigma^2} +  \frac{\partial u_B}{\partial \sigma_A} 
\frac{\partial P}{\partial u_B} \right]_{\sigma_B=0}  + (1-r) 
\left[ \frac{\partial \sigma^2}{\partial \sigma_A} 
\frac{\partial P}{\partial \sigma^2} +  \frac{\partial u_B}{\partial \sigma_A} 
\frac{\partial P}{\partial u_B} \right]_{\sigma_B=1} = 0 
\label{eq:x27} 
\end{eqnarray} 
when $\sigma_A=1/2\rho$. The conditioning on $\sigma_B$ here applies 
to the derivatives of $\sigma^2$ and $u_B$, but the conditioning has no 
implication on the derivatives of 
$P$. The reason is because $P$ is a function of $\sigma^2$ and $u_B$, 
but for $\sigma_A=\sigma_A^*=1/2\rho$, we have $u_B=1/2\sqrt{1-\rho^2}$ 
irrespective of $\sigma_B=0$ or $\sigma_B=1$, and likewise, 
we have $\sigma^2=1/4\rho^2(1-\rho^2)$ irrespective of 
$\sigma_B=0$ or $\sigma_B=1$. With this in mind, solving (\ref{eq:x27}) for 
$r$ we deduce, after a short algebra, that 
\begin{eqnarray}
r^* = \frac{1}{2\rho^2} \left[ \frac{ (2\rho^2-1)\frac{\partial P}{\partial \sigma^2} 
+ \rho^2\sqrt{1-\rho^2} \frac{\partial P}{\partial u_B} }
{ \frac{\partial P}{\partial \sigma^2} 
+ \sqrt{1-\rho^2} \frac{\partial P}{\partial u_B} }\right] \, , 
\label{eq:x28} 
\end{eqnarray}
where partial derivatives of $P$ are understood to be evaluated at $\sigma_A
=\sigma_A^*$. 

Next, making use of the identity $(1-p)\,n(d^-) = p\,n(d^+)\,\re^{-\alpha u_B}$ 
we deduce that 
\begin{eqnarray}
\frac{\partial P}{\partial \sigma^2} = p \, n(d^+) \left[ \frac{\partial d^+}{\partial 
\sigma^2}+\re^{-\alpha u_B} \frac{\partial d^-}{\partial \sigma^2} \right] 
\quad {\rm and} \quad 
\frac{\partial P}{\partial u_B} = p \, n(d^+) \left[ \frac{\partial d^+}{\partial 
u_B}+\re^{-\alpha u_B} \frac{\partial d^-}{\partial u_B} \right] \, .
\end{eqnarray} 
Hence the term $p \, n(d^+)$ drops out in (\ref{eq:x28}). A calculation shows 
that 
\begin{eqnarray}
\frac{\partial d^\pm}{\partial \sigma^2} = \frac{1}{2\sigma^2}\left[ -d^\pm 
\pm\sigma\sqrt{T} \right] \quad {\rm and} \quad 
\frac{\partial d^\pm}{\partial u_B} = -\frac{\alpha}{\sigma\sqrt{T}} \, ,
\end{eqnarray} 
from which it follows that 
\begin{eqnarray}
\frac{\partial P/\partial\sigma^2}{\partial P / \partial u_B} = 
\frac{1}{2\sigma^2\alpha} (\Lambda-\alpha u_B) - \frac{T}{4\alpha} \, 
\tanh\left( \half\alpha u_B\right) \, . 
\end{eqnarray} 
Substituting these in (\ref{eq:28}) we finally deduce that 
\begin{eqnarray}
r^* = \frac{1}{2\rho^2} \left[ \frac{(2\rho^2-1)\left[ 
\frac{\Lambda-\alpha u_B}{2\sigma^2 T} 
- \frac{1}{4} \tanh\left( \frac{1}{2}\alpha u_B\right) \right]+ \rho^2} 
{\frac{\Lambda-\alpha u_B}{2\sigma^2 T} 
- \frac{1}{4} \tanh\left( \frac{1}{2}\alpha u_B\right) + 1}  \right] \, , 
\label{eq:x32}
\end{eqnarray} 
where we recall that $\alpha=T/\sqrt{1-\rho^2}$, and where 
$u_B=1/2\sqrt{1-\rho^2}$ and $\sigma^2=1/4\rho^2(1-\rho^2)$ are the optimal 
parameter values in (\ref{eq:x32}). Substituting these in (\ref{eq:x32}) we 
obtain (\ref{eq:z32}). 
For a mixed strategy Nash equilibrium to be 
admissible we require $r^*\in[0,1]$. To determine the parameter rage for which 
the condition $r^*\in[0,1]$ holds, define 
\begin{eqnarray}
Z = \frac{\Lambda-\alpha u_B}{2\sigma^2 T} 
- \frac{1}{4} \tanh\left( \frac{1}{2}\alpha u_B\right) \, . 
\end{eqnarray}  
Then $r^*\in[0,1]$ if (a) $\rho\geq1/\sqrt{2}$ and either $Z\geq-1$or 
$Z<\rho^2/(1-2\rho^2)$, or if (b) $\rho<1/\sqrt{2}$ and 
$-1<Z<\rho^2/(1-2\rho^2)$. These conditions imply that $r^*\in[0,1]$ if 
$\rho\geq1/\sqrt{2}$ and either (\ref{eq:z20}) or (\ref{eq:z21}) holds, or 
if $1/2\leq\rho<1/\sqrt{2}$ and (\ref{eq:z22}) holds. 
Outside of these parameter ranges, no mixed strategy exists. In this case, 
because $u_B|_{\sigma_B=1}>u_B|_{\sigma_B=0}$ for all $\sigma_A\in[0,1]$, 
player B will choose $\sigma_B^*=1$ to maximise the impact of disinformation 
while player A will choose $\sigma_A^*=1$ to minimise its impact. 
\hfill $\Box$ 

The game of disinformation therefore is rather intricate and subtle even in the 
simplest model setup. The matter is further complicated by the fact that if player 
A let $\sigma_A$ approach the indifferent $\sigma_A^*$ from below, then 
$\sigma_B^*=0$, but if the value exceeds $\sigma_A^*$ then player B must 
choose $\sigma_B^*=1$, and this may also flip the sign of $\mu^*$. It would 
be of interest therefore to further explore different scenario analysis, especially 
in view of its importance in potential practical applications, for example, in a 
``game'' between a government (player A) and spreaders of disinformation that 
are harmful to the society (player B).

\section{Other open problems in information games}
\label{sec:6} 

There are other generalisations of the information game one can consider. 
One such generalisation is to study a game involving more than two players. 
For example, if there are three players with three information sources, their 
combined information flow rate $\sigma$ becomes a little more complicated 
\cite{BY2}. Writing $\sigma_A=\sigma_1$, $\sigma_B=\sigma_2$, and 
$\sigma_C=\sigma_3$ for the three control parameters, the overall information 
flow rate is determined by 
\begin{eqnarray}
\sigma^2 = \sum_{i,j=1}^3 \sigma_i \rho_{ij}^{-1} \sigma_j \, , 
\end{eqnarray}
where 
\begin{eqnarray}
\rho_{ij}^{-1} = \Delta^{-1} \left( \begin{array}{ccc} 
1-\rho_{23}^2 & \rho_{23}\rho_{31}-\rho_{12} & \rho_{12}\rho_{23}-\rho_{31} \\ 
\rho_{23}\rho_{31}-\rho_{12} & 1-\rho_{31}^2 & \rho_{31}\rho_{12}-\rho_{23} \\ 
\rho_{12}\rho_{23}-\rho_{31} & \rho_{31}\rho_{12}-\rho_{23} & 1-\rho_{12}^2 
\end{array} \right) \, , 
\end{eqnarray} 
with $\Delta=1+2\rho_{12}\rho_{23}\rho_{31}-\rho_{12}^2-
\rho_{23}^2-\rho_{31}^2$, is the inverse correlation matrix of the three noise 
terms $\epsilon_t^A$, $\epsilon_t^B$, and $\epsilon_t^C$. Thus the overall 
$\sigma^2$ remains a quadratic form of the individual control variables, but 
the way in which the probability of one of the choices dominating the other 
two at future time $T$, for example the winning probability of one of the 
candidates in an election, will depend on $\sigma$ in a nontrivial way \cite{BY1}. 
It would be of interest to explore whether an equilibrium solution can be found 
in closed form in a three-player information game. 

Another generalisation of a two-player information game concerns a situation 
in which the players must choose entire temporal information flow rates over 
the time period $[0,T]$. In this case, the two information sources take the 
form 
\begin{eqnarray}
\xi_t^A=X \int_0^t \sigma_s^A \rd s + \epsilon_t^A \quad {\rm and} \quad 
\xi_t^B=X \int_0^t \sigma_s^B \rd s + \epsilon_t^B \, . 
\label{eq:26}
\end{eqnarray} 
In such a time-dependent scenario, we have the following: 

\vspace{0.2cm}
\noindent 
{\bf Proposition 3}. \textit{
When the two information sources of {\rm (\ref{eq:26})} are combined, the 
resulting information can be expressed in the form 
\begin{eqnarray}
\xi_t = X \int_0^t \sigma_s \, \rd s + \epsilon_t \, , 
\label{eq:27}
\end{eqnarray}
where 
\begin{eqnarray}
\sigma_t = \frac{1}{\sqrt{1-\rho^2}} \, 
\frac{\sigma_t^A\int_0^t \sigma_s^A \rd s + \sigma_t^B\int_0^t \sigma_s^B \rd s
-\rho 
\left( \sigma_t^A\int_0^t \sigma_s^B \rd s + \sigma_t^B\int_0^t \sigma_s^A \rd s
\right)}{\sqrt{\left(\int_0^t \sigma_s^A \rd s\right)^2 - 2\rho 
\left(\int_0^t \sigma_s^A \rd s\right)\left(\int_0^t \sigma_s^B \rd s\right) + 
\left(\int_0^t \sigma_s^B \rd s\right)^2
\label{eq:28}
}}
\end{eqnarray}
is the combined information flow rate, with $\rho$ the correlation 
between $\epsilon_t^A$ and $\epsilon_t^B$, and 
\begin{eqnarray} 
\epsilon_t = \frac{1}{\sqrt{1-\rho^2}} \, 
\frac{\left(\int_0^t \sigma_s^A \rd s-\rho \int_0^t \sigma_s^B \rd s\right)\epsilon_t^A
+ \left(\int_0^t \sigma_s^B \rd s-\rho \int_0^t \sigma_s^A \rd s\right)\epsilon_t^B}
{\sqrt{\left(\int_0^t \sigma_s^A \rd s\right)^2 - 2\rho 
\left(\int_0^t \sigma_s^A \rd s\right)\left(\int_0^t \sigma_s^B \rd s\right) + 
\left(\int_0^t \sigma_s^B \rd s\right)^2
\label{eq:29}
}}
\end{eqnarray}
is a standard Brownian motion, independent of $X$. 
}
\vspace{0.2cm}

\noindent 
{\it Proof}. We begin by defining a Brownian motion ${\bar\epsilon}_t$, 
independent of $\epsilon_t^A$, such that we have $\epsilon^B_t = \rho 
\epsilon_t^A + \sqrt{1-\rho^2} \, {\bar\epsilon}_t$, and analogously define 
an information flow ${\bar\xi}_t$ by the relation $\xi^B_t = \rho 
\xi_t^A + \sqrt{1-\rho^2} \, {\bar\xi}_t$. Then we have 
\begin{eqnarray}
{\bar\xi}_t = \frac{\xi_t^B-\rho\xi_t^A}{\sqrt{1-\rho^2}} = 
X \int_0^t \frac{\sigma_s^B-\rho\sigma_s^A}{\sqrt{1-\rho^2}} \, \rd s + 
{\bar\epsilon}_t \equiv X\int_0^t {\bar\sigma}_s \, \rd s + {\bar\epsilon}_t \, .
\label{eq:30}
\end{eqnarray}
Next, define a ``pure-noise'' process $\{\delta_t\}$, independent of $X$, 
according to 
\begin{eqnarray}
\delta_t = \frac{\xi_t^A}{\int_0^t \sigma_s^A \rd s} - 
\frac{{\bar\xi}_t}{\int_0^t {\bar\sigma}_s \rd s} = 
\frac{\epsilon_t^A}{\int_0^t \sigma_s^A \rd s} - 
\frac{{\bar\epsilon}_t}{\int_0^t {\bar\sigma}_s \rd s} \, . 
\end{eqnarray} 
Our objective now is to identify a standard Brownian motion $\{\epsilon_t\}$ 
that is independent of $\{\delta_t\}$. Writing $\epsilon_t=\alpha_t \epsilon_t^A 
+ \beta_t \epsilon_t^B$, the two unknowns $\{\alpha_t\}$ and $\{\beta_t\}$ can 
be determined by the two conditions ${\mathbb E}[\epsilon_t\delta_t]=0$ and 
${\mathbb E}[\epsilon_t^2]=t$. Working out these unknowns we obtain 
$\{\epsilon_t\}$ of (\ref{eq:29}). In terms of $\{\alpha_t\}$ and $\{\beta_t\}$ 
thus obtained in (\ref{eq:29}), we define $\xi_t=\alpha_t \xi_t^A + \beta_t 
\xi_t^B$, and set $\{\sigma_t\}$ by the relation 
\begin{eqnarray}
\int_0^t \sigma_s \rd s = \alpha_t \int_0^t \sigma_s^A \rd s + 
\beta_t \int_0^t \sigma_s^B \rd s\, . 
\end{eqnarray}
Taking the derivative, we find that two of the terms cancel and we obtain 
$\sigma_t = \alpha_t \sigma_t^A + \beta_t \sigma_t^B$, which gives 
(\ref{eq:28}). It should be evident that information contained jointly in 
$\{\xi_t^A\}$ and $\{\xi_t^B\}$ is equivalent to information contained jointly in 
$\{\xi_t\}$ and $\{\delta_t\}$. However, the pure noise process $\{\delta_t\}$ is 
independent of $X$, so it drops out in calculation for the conditional probability 
law of $X$. It follows that the two information sources (\ref{eq:26}) can be 
combined into a single effective information source (\ref{eq:27}). 
\hfill $\Box$

In the case of time-dependent strategies $\sigma_t^A$ and $\sigma_t^B$, 
the expression for the initial likelihood ${\mathbb P}\left( \pi_T> K \right)$ 
remains to be given by (\ref{eq:8}), but $d^\pm$ now takes the form 
\begin{eqnarray}
d^\pm =  \frac{ \log\left[ \frac{p(1-K)}{K(1-p)}\right] \pm \frac{1}{2} 
\int_0^T \sigma_s^2 \rd s} {\sqrt{\int_0^T \sigma_s^2 \rd s}} \, .
\label{eq:33} 
\end{eqnarray}
Therefore, a time-dependent game of information reduces to a simpler problem 
of either maximising or minimising $\int_0^T \sigma_s^2 \rd s$, where 
$\{\sigma_t\}$ is given by (\ref{eq:28}) in terms of the two control variables 
$\sigma_t^A$ and $\sigma_t^B$. Hence we arrive at not merely a two-person 
infinite game, but rather a two-person functional game, where optimisation is 
over functional degrees of freedom.  

In summary, I have introduced the concept of information game, and showed 
that in the simplest model setup the game reduces to a more familiar case in 
standard game theory. I also considered a game of disinformation, and 
provided an analysis to show the existence of optimal equilibrium strategies. 
I then introduced two concrete generalisations of a simple game of information 
--- readers may come up with other generalisations and applications to explore. 
In view of the widespread prevalence of harmful disinformation in today's 
society, such explorations will no doubt be valuable to counter their impacts.

\end{document}